\documentclass[prl,twocolumn]{revtex4}
\usepackage{amssymb}

\usepackage{graphicx}
\usepackage{subfig}

\newcommand{\hsp}{\hspace{ 0.5 cm}}
\newcommand\beq{\begin{equation}}
\newcommand\eeq{\end{equation}}
\newcommand\bea{\begin{eqnarray}}
\newcommand\eea{\end{eqnarray}}

\newcommand\eps{\varepsilon}
\newcommand \ds [1]{ \displaystyle{ #1} }
\newcommand \nn{ \nonumber }

\usepackage{axodraw}
\usepackage[applemac]{inputenc}
\usepackage[francais]{babel}
\usepackage[T1]{fontenc}

\begin{document}

\title{The thermopower as a signature of quantum criticality in heavy fermions }
\date{\today}
\author{  Ki-Seok Kim$^{1}$ and C. Pépin$^{2,3}$}
\affiliation{$^1$ Asia Pacific Center for Theoretical Physics,  POSTECH, Hyoja-dong, Namgu, Pohang 790-784, Korea \\
$^2$Institut de Physique Théorique, CEA-Saclay, 91191 Gif-sur-Yvette, France\\
$^3$International Institute of Physics,
Universidade Federal do Rio Grande do Norte,
Rua Odilon Gomes de Lima, 1722
59078-400 Natal-RN,
Brazil\\
}

\begin{abstract}
We present a series of arguments showing that the Seebeck coefficient can be used as a decisive experiment to characterize the nature of the quantum critical point (QCP) in heavy fermion compounds. Being reactive almost exclusively to  the  presence of delocalized entropic carriers, the Seebeck coefficient  shows a drastic collapse at the Kondo breakdown QCP, as the reconstruction of the Fermi surface takes place.
In contrast, around a Spin Density Wave  (SDW) QCP,  the Seebeck coefficient is broadly symmetric.
We discuss the possibility of a change of  sign at the QCP, the characteristic variation  of $| S / T | $ with temperature and external parameter,
 as well as the capacity of  the Seebeck coefficient to distinguish between localized and itinerant anti-ferromagnetism (AF). Suggestions of experiments are given in the case of four non conventional compounds : YbRh$_2$Si$_2$,  Ce(Mn)In$_5$, CeCu$_{6-x}$Au$_x$ and URu$_2$Si$_2$.
\end{abstract}

\maketitle

\section{Introduction}

 In the mid-eighties, heavy fermion compounds were intensively studied for their heavy
 Fermi liquid  properties \cite{stewart,QimiaoFrank,MatthiasHilbert}.  Although  those dense rare earth lattices were made of big localized atoms supporting
 big magnetic moments, the low energy properties remained in the universality class of the Landau Fermi
 liquid theory of metals, with a characteristic low temperature saturation of the  Sommerfeld coefficient $\gamma = C/T $
 of the magnetic susceptibility  $ \chi_0 \sim Cst $ and a $T^2$-dependence of the corrections to the residual resistivity  $ \rho (T) - \rho_0 = A T ^2$. The effective mass of the Landau quasi-particles is strongly re-normalized,  up to a factor of $\sim 1000$ for UBe$_{13}$, but still the observed low temperature properties didn't depart from the Landau theory of metals.

 The situation changed drastically in the mid-nineties, with the observation that, under the application of  an external tuning parameter
 like chemical doping, pressure or magnetic field,  the specific heat coefficient doesn't saturate when the  temperature is lowered \cite{rosch}. This anomalous property was rapidly
  attributed to the presence of a QCP where the system orders anti-ferromagnetically at
  (theoretically) vanishing temperature. The strong quantum fluctuations induced at the vicinity of a zero temperature phase
  transition were suggested to be responsible for this  violation of the Landau theory of metals. Rapidly other properties
 were shown not to follow the universal Landau paradigm. In many compounds the resistivity is linear,
 or quasi-linear in temperature over two or three decades in energy \cite{stewart,YRSres}, the magnetic susceptibility shows some anomalous exponents with temperature like in CeCu$_{5.9}$Au$_{0.1}$\cite{schroeder}.  As the tuning parameter evolves from the heavy  Fermi liquid phase towards the QCP, the A-coefficient of the $T^2$-resistivity
 shows a divergent trend with respect to the tuning parameter\cite{YRSupturn}. The effective band mass, shown by de Haas van Alphen experiments is strongly re-normalized at the approach of a QCP\cite{onukiCeRhIn5}, which is a remarkable fact since the renormalization of the band mass is essentially due to elastic scattering processes.
 All over the years, an average of 20 compounds were found to have anomalous physical properties, when fine tuning with an external parameter was performed. Those   findings are well summarized in various review articles \cite{PiersDie,QimiaoFrank,MatthiasHilbert}  and we refer the reader to them for further details.

 In heavy fermion compounds the study of the thermo-power started three decades ago \cite{jaccard1}  and a few systematic features were already clarified. The high temperature thermopower is typically large due to the interplay of incoherent spin fluctuations and crystal field effects.  Like most of the thermodynamic and transport properties, the termopower shows  a maximum corresponding to the  bandwidth of the  f-electrons. This scale is sometimes refered to  as the lattice coherence scale $T_0$ \cite{pines-two-scales}. The sign of the Seebeck coefficient has been shown to depend crucially on the position of the f-resonance level with respect to the Fermi surface. In the Ce (4f$^1$)-series the f-resonance sits above the Fermi level,which leads to a positive Seebeck coefficient, while in the case of the Yb (4f$^{13}$)-based compounds, the f-level is below the Fermi level which leads to a negative Seebeck coeffcicient. The case of the U (4f$^2$)-based compounds is more controversial, since some compounds are compensate metals with very low carrier density, like URu$_2$Si$_2$ and have a negative Seebeck coefficient \cite{sakurai} while for UPt$_3$ no detectable signal has been observed \cite{tholence}. For UBe$_{13}$ , $S/T$ is strong and negative\cite{jaccard2} while for UPd$_2$Al$_3$ and UNi$_2$Al$_3$\cite{grauel} it is small and positive, revealing the complexity of the f-electron structure in U-based compounds.

 Quite remarkably, at low temperature in heavy fermions, the Seebeck coeffcient divided by the temeprature $S/T$ was shown to form a constant ratio with the Sommerferld specific heat coefficient $ \gamma $ \cite{kjf}
 \beq
 q = \frac{S}{T} \ \frac{N_{Av} e }{\gamma} \label{eqn1} \ ,  \eeq where $N_{Av}$ is the Avogadro number and $e$ the electron's charge.This ratio is close to $±1$ for the majority of heavy fermion compounds. This quasi-universal behavior was explained by the observation that
 although many bands are present in a typical  band structure of heavy fermions, the Seebeck coefficient is mostly sensitive to the position of the heaviest band, namely the one with the biggest f- character\cite{miyake1}. Since the Sommerfeld ratio is precisely sensitive to the  heaviest band as well, a quasi-universal behavior is to be expected. The formula (\ref{eqn1}) tells us that, in the Fermi liquid regime, the thermopower probes the specific heat per electron. This ratio can be compared with other quasi-universal ratios studied in heavy fermion systems.  The Wilson ration $ \chi/\gamma$ \cite{wilson} of the magnetic susceptibility to the Sommerfeld coefficient and the Kadowaki-Wood ratio $A / \gamma^2$\cite{kw}  of the  coefficient A of the $T^{2}$ resistivity in metals show some universal ratio insensitive to the mass renormalization in  of the heavy fermi liquid.

 In simple metals, the thermoelectric effects are very sensitive to the type of scattering involved. In addition to the diffusion Seebeck coefficient, the electron-phonon interaction produces a phonon-drag component which dominates the behavior
  in many metals \cite{barnard}. In the presence of various types of scattering the thermopower is the sum of the contribution of each scattering process, weighted by the resistivity,  a rule reminiscent of the Matthiessen  rule for the addition of resistivity, referred to as the Nordheim-Gorter rule
  \beq
  S =\frac{ \sum_i  \rho_i S_i}{\sum_i \rho_i} \  . \eeq In the case of multi-bands systems, the Mott \cite{mott} rule applies where the Seebeck coefficient for each band is weighted by the conductivity
  \beq S =\frac{ \sum_i  \sigma_i S_i}{\sum_i \sigma_i} \  . \eeq

  Little is known about the Seebeck coefficient close to a QCP. Preliminary studies for CeCu$_{6-x}$Au$_x$ \cite{lohneysen}  and
  Ce( Ni$_{1-x}$Pd$_x$)$_2$Ge$_2$\cite{kuwai} show that the  presence of a QCP modifies  low temperature dependence of the
  Seebeck coefficient. Two recent studies under magnetic field show some striking similarity between thermoelectric effects in
  CeCoIn$_5$ \cite{kamran2} and URu$_2$Si$_2$ \cite{kamran3}. In particular both system show a pronounced anisotropy in their
  thermoelectric response.  Lastly, a recent experiment on YbRh$_2$Si$_2$ under a small magnetic field shows some drastic variations of the magnitude of the Seebeck coefficient on both sides of the QCP\cite{hartmann}.

  Even fewer theoretical studies are available \cite{miyake1,pruschke,zlatic}. In the case of the Spin Density Wave (SDW)  QCP, the authors of Ref.\cite{indranil}  have shown that at the QCP  $S/T$ has the same variation with temperature as the Sommerfeld coefficient $\gamma(T)$.The low temperature correlation between  $S/T$ and $\gamma$  survives close to a QCP.

  In this discussion paper, we want to address the relevance of thermoelectric properties close to the Kondo breakdown QCP.
  In the next section we give an overview of the Kondo breakdown theory and explain  why the Seebeck coefficient might be the best probe to characterize the Kondo breakdown QCP. We make as well some distinctions between the SDW scenario and the Kondo breakdown, which can lead to experimental discrimination between the two QCP.
  In the next section, we review  the unconventional properties of  QCPs in four heavy fermion compounds
  and suggest useful thermoelectric experiments susceptible to unravel the true nature of the  QCP.

  \section{Unconventional QCPs: the Kondo breakdown  model}

   In this section we review  the  various QCPs that have been suggested
   to explain the very unconventional behavior observed in heavy fermions.
 Heavy fermions are heavy metals made of big magnetic atoms interacting hybridized to a bath of conduction electrons.
 Many compounds exhibit magnetic phases,  anti-ferromagnetic (AF) or frustrated. It was natural to attribute the anomalous
 properties of those compounds to the proximity to a magnetic phase transition at $T=0$. At this QCP,  the Fermi surface is destabilized by
 spin density waves. This scenario is also called the SDW  theory. It has been derived by Hertz \cite{hertz} and revived by Millis \cite{millis}.  At the heart of this theory is the itinerant character of  conduction electrons. When a bosonic mode of the type of a SDW interacts with conduction electrons, the particle-hole continuum  produces Landau damping $ - i \omega / q $, where q is the  modulus of the scattering vector. If  the QCP sits at the brink of uniform ordering, like for example in the case of a ferromagnet  we are in the regime where $ q \rightarrow 0 $ and $ |\omega | \leq q $.

 For incommensurate or AF order though,  the ordering wave vector is finite and  the damping takes the form $ -i  \omega / Q^* $, where $Q^*$ is the modulus of the ordering wave vector. In this case the spin susceptibility  in the vicinity of the QCP writes
 \beq  D_\chi^{-1} (Q, \omega) \propto  \frac{ - \gamma i \omega}{ Q^*}  + q^2 + \xi^{-2} \ , \eeq where $\gamma =  2 \pi m $  and $ m$ is the band mass of the conduction electrons. $\xi$ is the correlation length which depends both on the temperature and on the distance from the QCP; at the QCP $\xi \rightarrow \infty $. We see that  in this model, the fluctuations in the imaginary time, also called the quantum fluctuations, scale like $\omega \sim q^2$ which defines the dynamical exponent $z=2$ \cite{note1}. The treatment of the Hertz-Millis theory requires to integrate the fermions out of the partition function, which is an uncontrolled operation. A better treatment is given by writing a set of self-consistent equations for the polarization and the self-energy and using the Migdal theorem to neglect vertices. This amounts to  performing an Eliashberg treatment of this theory \cite{chub1,rech-chub}. Those two techniques give the same result, and  in the absence of a reliable bosonization  of the SDW model \cite{efetov1}, they constitute the state of the art.

\begin{figure*}[htbp]
  \begin{center}
    \leavevmode
    \subfloat[SDW phase diagram for itinerant QCP]{%
      \label{fig1a}
      \includegraphics[height=5cm]{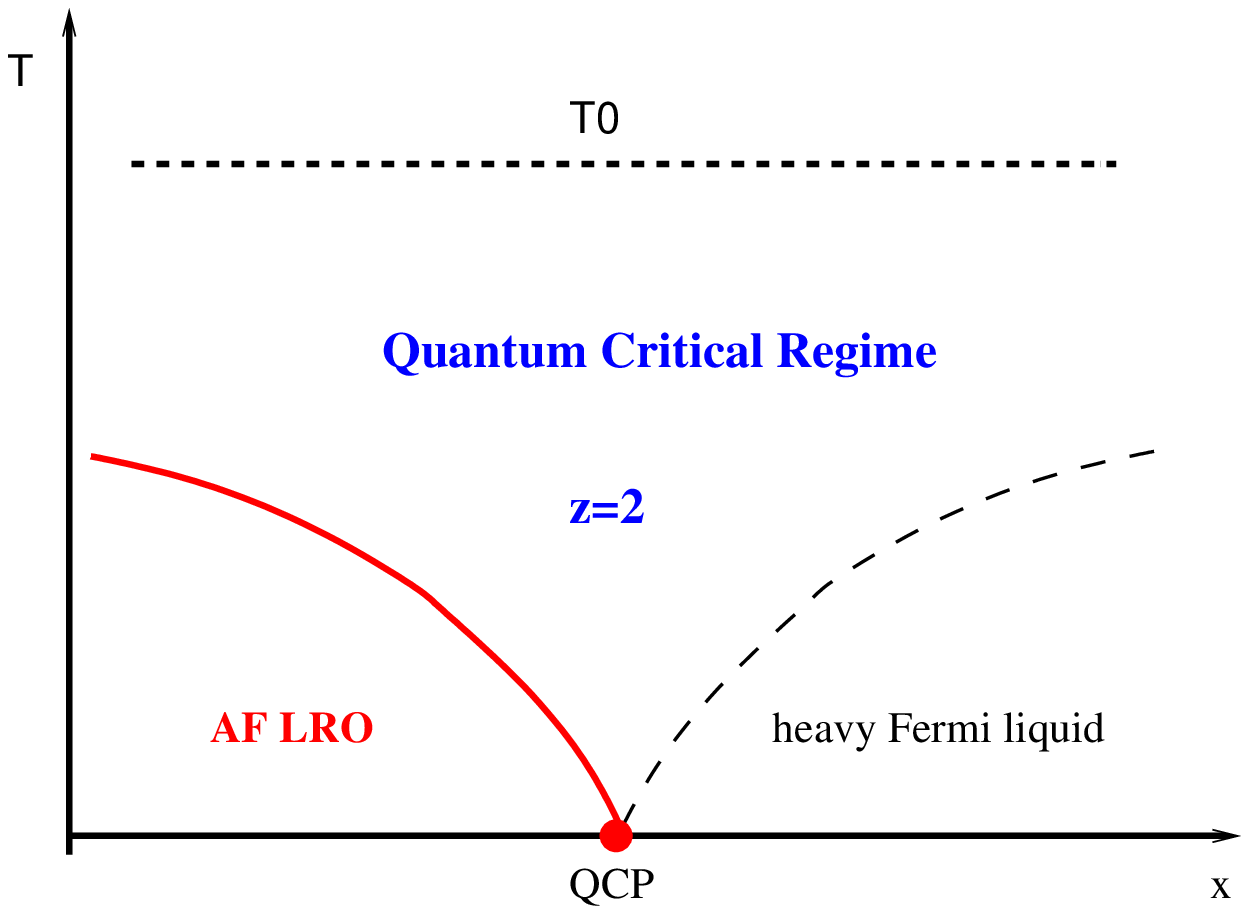}}
    \hspace{1 cm}
    \subfloat[Kondo breakdown phase diagram for heavy fermions]{%
      \label{fig1b}
      \includegraphics[height=5cm]{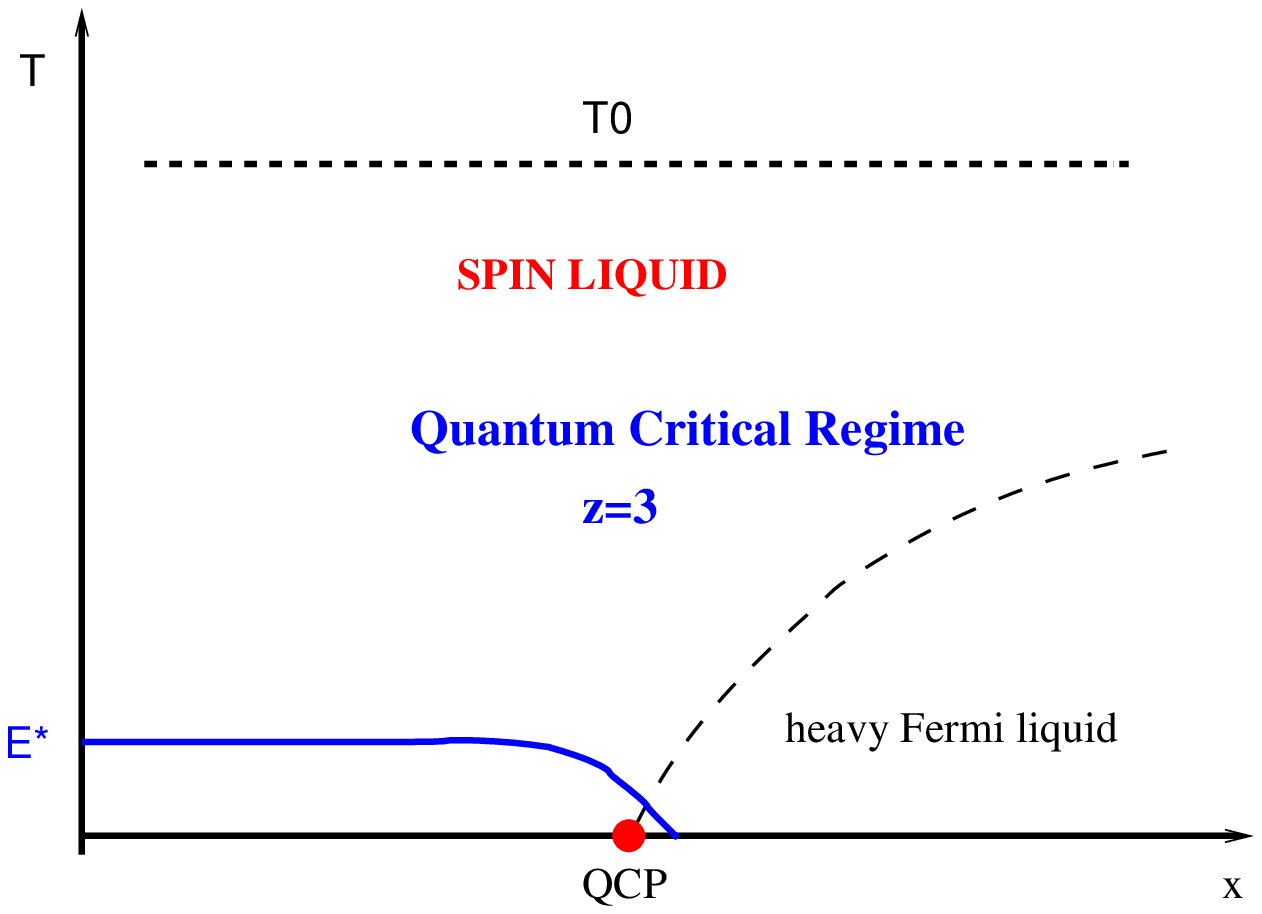}}
    \caption{ This figure compares the phase diagram of the Kondo Breakdown QCP with the one of the SDW model at and AF phase transition. In the SDW model the f-electrons remain itinerant while crossing the QCP while in the Kondo Breakdown,
    the f-electrons localize below $E^*$  on the left-hand side of the phase diagram. In the  Kondo Breakdown, the scale $T_0$ marks
    the onset of a spin liquid, where entropy is quenched by entanglement of the f-moments with no long range order; in the SDW $T_0$
    can be any scale associated with the mean-field formation of Kondo singlets. A crucial difference between the two QCPs is that the quantum critical regimes have different dynamical exponents, with $z=2$ for the SDW and $z=3$ for the Kondo breakdown. That means that experimental observable have different exponents in this regime. The scale $E^*$ is typical of the Kondo breakdown, it marks the end of the $z=3$ regime.  The regime below $E^*$ has Fermi liquid characteristics. AF long range order can occur at the vicinity of the Kondo Breakdown QCP but is directly tight with the zero temperature phase transition. It has been omitted in this figure. Note that the scale $E^*$ made to stop at the  boundary of the heavy Fermi liquid regime, since this scale is difficult to observe in this regime.}
    \label{fig1}
  \end{center}
\end{figure*}

 The results  obtained from the SDW model are summarized in the left panel in Fig.\ref{fig1a} and in Table \ref{table1}. In the quantum critical regime, the Sommerfeld coefficient diverges logarithmically at low temperature in dimension two but not in dimension three. Its generic scaling with the temperature  goes like  $  \gamma \sim T^{(d-2)/2}$. The corrections to the electrical  resistivity vary like $\rho- \rho_0 \sim T^{d/2} $. This power law  has to be understood as a correction to the residual resistivity. It is valid when  $ \rho -\rho_0 \ll \rho_0 $. The static staggered spin susceptibility varies like $T^{d/2}$. The Seebeck coefficient divided by the temperature \cite{indranil} varies like the Sommerfeld coefficient, as $T^{(d-2)/2}$. When crossing the QCP by decreasing or increasing the external parameter ``x'' (here ``x'' represents pressure, doping or a small magnetic field) a doubling of the Brillouin zone is observed, but with conservation of the Luttinger theorem; the number of electronic carriers is conserved.

In contrast with the SDW model, the Kondo Breakdown QCP is not associated with  a $T=0$ magnetic phase transition but with the localization of the f-electrons under very strong onsite Coulomb potential $U$.  When the QCP is crossed from the right to the left by varying ``x'', a complete reconfiguration of the Fermi surface is observed; the Fermi surface of the f-electrons becomes hot and on the left side of the phase diagram the f-electrons don't participate to the transport. 
This transition has been dubbed in some works selective Mott transition, the word selective referring to the localization of the f-electron while the conduction electrons remain itinerant. The $T=0$ phase transition in this case is a de-confinement transition for the heavy electron.

 At the QCP the heavy electron splits into three parts,
(i) the conduction electrons,
 (ii) the spinons ($\chi_\sigma$, $\chi^\dagger_\sigma$) carrying spin and
 (iii) the holons ($b, b^\dagger$) carrying charge,  describing the break-up of the f-electron at the Mott localization. In a field theory language this transition is described as a condensation of the holon-operator $b= b^\dagger= b_0$, within a spontaneous symmetry breaking also called Anderson-Higgs transition. Fictitious gauge fields are generated to
sustain the U(1) gauge symmetry. The whole description with spinons and holons can be understood as a field-theoretical way of tracking the Mott transition, analogous to what was implemented for the single band Hubbard model in the early days of the cuprate superconductors\cite{patrick1}.

One of the main differences between the SWD model and the Kondo Breakdown is that the Kondo Breakdown has a  $z=3$ quantum critical regime (instead of the $z=2$ quantum critical regime of the AF SDW model) \cite{note5}.  In this regime the typical form of the holon propagator reads \cite{note2}
\beq  D_b^{-1} (q, \omega) \propto  \frac{ - \gamma i \omega}{ \alpha q}  + q^2 + \xi^{-2} \ , \eeq where $\alpha$ is a dimensionless number much smaller than one, which represents the ratio between the $\chi$-spinons  and the c-electrons bandwidths.
As a result, the critical exponents are different from the ones of the  SDW thoery. The Sommerfeld coefficient varies like $\gamma \simeq T^{(d-3)/3}$. The resistivity varies like $\rho - \rho_0 \simeq T^{d/3}$  and the Seebeck coefficient over the temperature varies like $S/T \simeq T^{(d-3)/3}$.  It is worth noticing  that in dimension three, the resistivity is quasi-linear in temperature, with $ \Delta \rho \simeq T Log ( T/ E^*) $. The quasi-linear  temperature exponent is not  correction to the residual resistivity, but a robust exponent  due to  the specificity of the Kondo Breakdown to have two kinds of particles, light conduction electrons and almost localized spinons. The light conduction electrons scatter through the  local network of spinons, the scattering process involves the $z=3$ critical bosons, producing a transport exponent quasi-linear in temperature \cite{paul2}. The critical exponents for the SDW and the Kondo Breakdown models are compared in Table\ref{table1}.
\begin{table}[htdp]
\begin{tabular}{|c|c|c|}
\hline
 \hspace{1.5 cm}   &  \hspace{1cm} SDW  \hspace{1 cm} &\hsp KBreakdown \hsp  \\
  \hline
   $\Delta \rho  $  & $  T^{d/2} $  &  $ T^{d/3} $  \\
  $\gamma $ & $ T^{(d-2)/2}  $ & $ T ^{(d-3)/3} $  \\
 $ S/T $&$  T^{( d-2)/2} $ & $ T^{(d-3)/3} $ \\
  \hline
  \end{tabular}
\caption{comparison  of the critical exponents for the SDW ($z=2$) and the Kondo Breakdown ($z=3$) models, for the resistivity, the Sommerfeld coefficient and the ratio of the Seebeck coefficient with the temperature.}
\label{table1}
\end{table} The Kondo Breakdown model also differs from the SDW theory from because of the emergence of an additional scale -called $E^*$-, at the QCP.The  scale $E^*$, in this model, is due to the presence of  two types of fermions, the $\chi$ spinons and the conduction electrons. The two corresponding Fermi surfaces are not necessarily close to each other. In the case where they are centered, the mismatch $q^* = | k_F^\chi- k_F^c |$ between the Fermi wave vectors  produces an additional energy scale
\beq  \label{estar}
E^* \simeq 0.1 (q^*/ k_F^c )^3 T_0  \ , \eeq  where $T_0$ is the scale above which the entropy $R \ln 2 $ is quenched. In the case where the two Fermi surfaces are not centered, the holons condense at finite $q_0$ in order to re-center them \cite{note5}.
Note that the power 3 exponent  in Eqn.(\ref{estar}) ensures that for quite a number of compounds $E^*$ is small; typically $E^*_{min} \leq E^* \leq E^*_{max}$ with $E^*_{max} \simeq 250 m K $ for $T_0 \simeq 20 K$ and $ q^*/k_F \simeq 0.5$ and $E^*_{min} \simeq 2. 10^{-6} K$ for $T_0 \simeq 20 K $ and $ q^*/k_F \simeq 1. 10^{-2}$.

The scale $E^*$ is a key feature of the Kondo breakdown QCP. Below $E^*$ the particle-hole continuum is gapped out,
 hence the order parameter reduces to a free boson mode below the gap with the dispersion $  D_b^{-1} (q, \omega) \propto \omega  + q^2 + \xi^{-2} $. This mode doesn't lead  to any appreciable contribution to the thermodynamics and transport, and the  regime below $E^*$ can be characterized by small corrections to the Fermi liquid theory \cite{notestar}.
The reconfiguration of the Fermi surface at the QCP can be found in Ref.\cite{senthil1}. The multi-scale character of the QCP, as well as the $z=3$ regime can be found in the  work of Ref.\cite{paul2}. The Kondo breakdown QCP has already been the object some scrutiny by various groups \cite{tous}. In particular,  two DMFT studies are now confirming its existence \cite{deleo,ferrero}. The Kondo breakdown can be described through an effective low energy field theoretical Lagrangian, which enables refine the theoretical predictions. The most complete treatment to date, however, still relies on an Eliashberg theory where the vertices are neglected  and the self-energies retained.
 At this stage of development, the theory suffers form the fact that the localized spinons are described within a fermionic representation of the spin (Abrikosov pseudo-fermions). Hence the  properties  associated with the entropy of the localized spins are poorly described. We expect however, that the model gives a correct description of the transport properties.

    Another scenario has been proposed in the literature to explain the anomalous properties observed in those compounds: the locally quantum critical scenario \cite{qimiao}. This theory is also based on a breakdown of the heavy Fermi liquid, and thus enter the generic class of `` Kondo Breakdown'' scenarios. However it leads to different results as the Kondo breakdown QCP and it is supported by a few assumptions that we believe will become experimentally testable in the near future.  The locally quantum critical point requires the presence of two dimensional spin fluctuations. It predicts some anomalous exponents in the spin susceptibility $\chi \sim T^{- 0.75}$ over a wide range of the Brillouin zone. Moreover, it is always  situated at the brink of a magnetic  $T=0$ phase transition. This property distinguishes it from the Kondo Breakdown QCP, which not directly correlated to  the occurrence of long range magnetic order.

\begin{figure*}[htbp]
\begin{center}
\includegraphics[width=14cm]{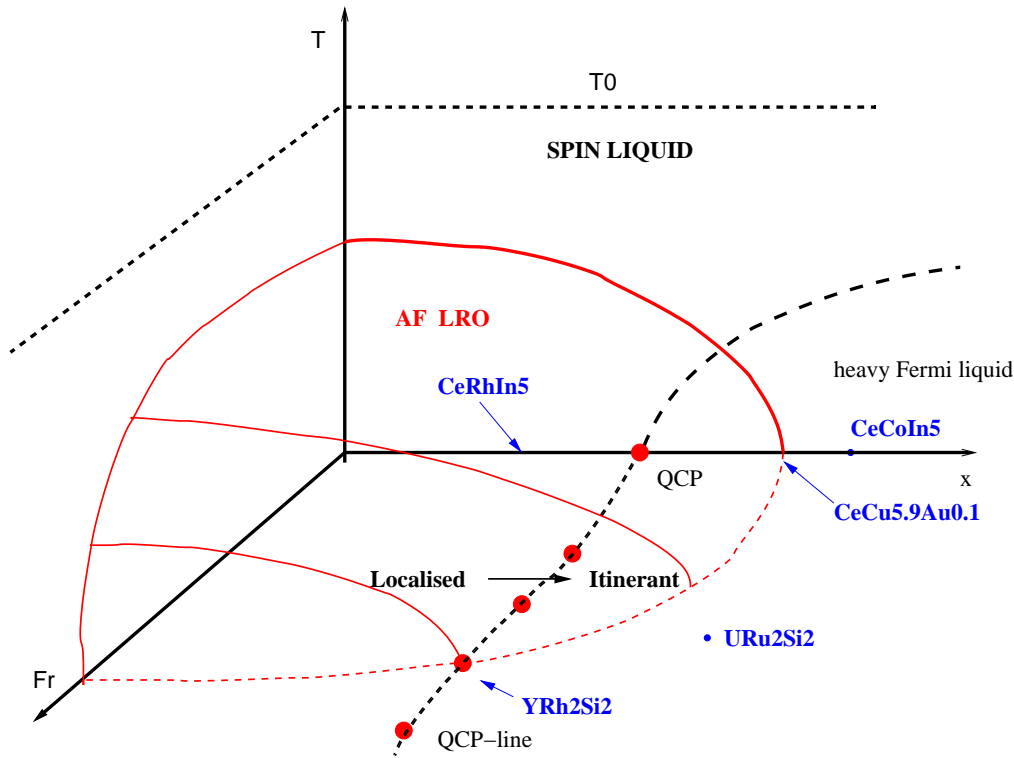}
\caption{Tentative phase diagram of the Kondo Breakdown QCP in the presence of magnetism. The third axis represented on this diagram is the axis of ``frustration''. It can be any external parameter which competes with the AF long range order. When the frustration parameter is strong enough, AF disappears, revealing the Kondo Breakdown QCP. Within this 3D phase diagram, one observed a line of Kondo Breakdown QCPs, which are uncorrelated with the magnetic order. The crossing  of the two critical lines  of AF LRO and Kondo Breakdown is accidental.  In the Kondo Breakdown theory, the compound YbRh$_2$Si$_2$ is situated at the crossings; CeRhIn$_5$ would be situated somewhere on the frustration axis, URu$_2$Si$_2$ would be deep in the heavy Fermi liquid phase (with a super-conducting instability at low temperatures) and CeCu$_{6-x}$Au$_x$ is located at the AF QCP of itinerant character. This phase diagram suggests that the Kondo Breakdown  QCP is a generic  feature of any heavy fermion phase diagram; it is a universal fixed point, of non magnetic character, whose
influence  on transport properties  dominates other scattering mechanisms in the quantum critical regime. Note that another phase diagram has been proposed \cite{qimiaonew} where the crossing of the Kondo breakdown line and the AF line has a finite width. }
\label{fig2}
\end{center}
\end{figure*}

    It is interesting to  find out what kind of phase diagram one obtains when the Kondo breakdown and  magnetism are treated together. At  the present moment the available theories don't allow us describe both phenomena together in a controlled way, but it is still interesting to consider the putative phase diagram one would obtain. The result is  presented in Fig \ref{fig2}, where an additional axis of frustration has been added to the system. Due to the Rudderman-Kittel-Kasuya-Yoshida (RKKY) interactions, frustration is naturally generated in the Kondo lattice, and it is  interesting to think that the combined effects of crystal fields and geometric frustration vary from compound to compound and lead to various magnitude and different structure for the AF magnetic order.  In the 3d-phase diagram, a line of Kondo Breakdown QCPs is crossing the AF long range order line (at $T=0$) at one point only,  and this crossing is accidental. The line of Kondo Breakdown QCPs separates
    a localized regime on the left  to an itinerant regime on the right.

The phase diagram of Fig \ref{fig2} presents some analogies  with the two fluids scenario for heavy fermions, which  has been proposed as a
generic mechanism for the formation of  the heavy Fermi liquid \cite{two_fluids}. In order to understand better the analogy, it is instructive to think about those compounds in terms of entropy.  At high temperature, heavy fermions are made of a lattice of fluctuating spins,  having a marginal and incoherent  interactions with the conduction electrons. As temperature is lowered, this entropy is quenched by spin-orrbit and crystal field coupling, but still, at intermediate temperature a finite entropy of the order of $R \ln 2$ per impurity site remains. As temperature is lowered further down, two routes open to quench the entropy. Either the impurity spins start to entangle together, quenching the entropy via the formation of a paramagnetic liquid which is called here a ``spin liquid'', or the entropy can be quenched by the formation of Kondo singlets, which for the lattice finally leads to the formation of the heavy Fermi liquid. Lastly at even lower temperatures, long range order of various kinds can achieve further quenching of the entropy. Mainly AF order or super-conductivity occur as a rule.
Those two routes for the quenching of the entropy - the formation of a spin liquid or the one of a heavy Fermi liquid- have been intuited  forty years ago by Doniach \cite{doniach}. He was the first to understand that the competition between short range magnetism and the formation of the Kondo singlet was an important key to understand the emerging phases in those compounds.  For the Kondo Breakdown theory the reasoning is similar; two forces compete at intermediate energy scales, a  frustrated magnetic force leading to the formation of the spin liquid and
and the Kondo interaction leading to the formation of the heavy Fermi liquid.
In the phase diagram in Fig \ref{fig2}, $T_0$  is the temperature  above which the entropy $R \ln 2$ is released. In the Kondo Breakdown theory,  $T_0$ is understood as the bandwidth of the f- spinons, and thus is associated with quenching of the entropy through the RKKY interactions, with formation of a spin liquid. The formation of the heavy Fermi liquid occurs further on the  right side of the phase diagram, where the conventional heavy fermion metals are located. For convenience, it is not represented in Fig \ref{fig2}.
It is interesting to notice that the two-fluids model of Ref. \cite{two_fluids} comes to the same conclusion,  that is in the compounds  for which anomalous transport and thermodynamic  properties have been observed-, the formation of the spin liquid occurs before the formation of the heavy Fermi liquid. Likewise in the vicinity of the Kondo Breakdown QCP, the  temperature $T_0$ is associated to the RKKY interactions rather than to the mean-field Kondo scale.

\section{Thermopower in the vicinity of a QCP}

\subsection{In the vicinity of the SDW QCP}

We now turn to the study of the thermopower in the vicinity of  a SDW QCP. The quantum critical regime is described in Ref. \cite{indranil} while  some insight about the saturation in the  zero temperature regime can be found in \cite{miyake1}. Here we summarize these two bodies of results and present a derivation of the thermopower in all the regimes around the QCP. We start with the definition of the thermopower as a ratio of two correlation functions
\beq
S= \frac{L_{12}}{e T L_{11}} \ , \eeq  where $L_{12}$ is the correlation function between the  heat current and the electrical current, and $L_{11}$ is the current-current correlation function defined as
\[ \begin{array}{cc}
 L_{11} = \lim_{\omega \rightarrow 0 } \frac{1}{ \omega} \ Im \int_0^\beta d \tau e^{i \omega \tau}  \left \langle  T_\tau {\bf j}(\tau)\cdot {\bf j} ( 0 ) \right  \rangle \  ,  \\ \\
 L_{12} = \lim_{\omega \rightarrow 0 } \frac{1}{ \omega} \ Im \int_0^\beta d \tau e^{i \omega \tau}  \left \langle  T_\tau {\bf j}_Q(\tau)\cdot {\bf j} ( 0 )  \right \rangle \ , \end{array}   \] where ${\bf j } _Q$ is the heat current and ${\bf j}$ the electric current. Those two operators can be put into the following form
 \beq \begin{array}{ll}
 L_{11}= \sum_{\bf p} v_{\bf p } ^2 \int_{- \infty} ^{ + \infty }  d \omega   \left ( - \frac{ \partial f} {\partial \omega} \right )  A^2( {\bf p}, \omega)  \ , \\ \\
 L_{12}= \sum_{\bf p} v_{\bf p } ^2 \int_{- \infty} ^{ + \infty }  d \omega   \left (-  \frac{ \partial f} {\partial \omega} \right )   \omega A^2( {\bf p}, \omega)  \ , \end{array} \label{eqn2} \eeq where $ {\bf v } _{p } =  \partial \epsilon_{\bf p}/ \partial {\bf p } $ is the velocity of the quasi-particles, that we  consider un-renormalized by the fluctuations, and $A({\bf p } , \omega) $  is the spectral function. We use here the notation of Ref.\cite{indranil} and define it as
 \[
 A({\bf p}, \omega) = \frac{\tau_{\bf p}^{-1}( \omega) }{ ( \omega/Z_\omega - \epsilon_{\bf p})^2 + \tau_{\bf p }^{-2}( \omega) }  \  . \] Here $Z_\omega$ is the quasi-particle  weight defined as
 \[ Z_\omega^{-1} =  1 - \frac{\partial Re \Sigma_c ( k_F , \omega) }{\partial \omega}  \]  and $\tau_{\bf p } ( \omega) $ is the transport scattering time, which includes both the effects of the impurities and the scattering through the fluctuations of the bosonic mode.  Another difference with Ref. \cite{indranil} is that   $ \tau_{\bf p }$ depends on  the position of ${\bf p} $ on the Fermi surface.
 We use the Mathiessen's rule for adding the resistivities to get
 \beq \label{tau}
 \tau_{\bf p} ^{-1} (\omega)    =  \tau^{-1}_{imp} ({\bf p },  \omega) + \tau ^{-1}_ {dyn}  ( {\bf p },\omega  ) \  . \eeq
 To simplify the discussion, we take  $ \tau^{-1}_{imp} ({\bf p },  \omega) = \tau_0^{-1}$ as a constant of ${\bf p }$ and  $ \omega$. The elastic scattering time $\tau_0$ encompasses for example the scattering through impurity centers. The effect of the fluctuations are described by $\tau^{-1}_ {dyn}  ( {\bf p },\omega ) = \tau^{-1}_{h} $ in the hot regions and $\tau^{-1}_ {fluct}  ( {\bf p },\omega ) = \tau^{-1}_{c} $  in the cold regions. Typically in the SDW theory the inelastic part of the  scattering time has the following form
 \beq \begin{array}{l} \label{taubis}
 \tau^{-1}_h \simeq A_h  \  T^{(d-2)/2} \\
 \tau^{-1}_c \simeq A_c \ T^2 \ , \\
 \end{array} \eeq where $A_h$ and $A_c$ are non universal constants.  $\tau^{-1}_c$ has the typical Fermi liquid  exponent while $\tau^{-1}_h$ has an anomalous exponent due to the scattering through the soft quantum modes present at the QCP.
 Details of the evaluation of $L_{11}$ and $L_{12}$ can be found  in the Appendix. The result is:
 \bea \label{eqn8} L_{11} & = &    \frac{ \pi v_F^2 \rho^*_0}{2}  \left [ \frac{ V_{h}}{\tau_0^{-1} + \tau^{-1}_{h} }  + \frac{ V_{c}}{\tau_0^{-1} + \tau^{-1}_{c} } \right ] \ ,
 \\ &  &  \nonumber \eea  where $\rho^* d\eps = \int_0^{+ \infty} p^2 dp / ( 2 \pi )^2 $ and $V_{h} $ (resp. $V_{c}$~) is the volume of the hot (resp. cold~) regions of the Fermi surface, satisfying $ V_{h} + V_{c}  = V_F$, the total volume. For an AF in $D=3$ where we take a spherical Fermi surface with hot lines at the angle $ \phi_0$ we get $V_{h} = \sin \phi_0 \ \Delta \phi ( T)  \sim \sqrt {T} $ and $V_{c} = 2 - \sin \phi_0  \ \Delta \phi ( T) $.  In the case  of two dimensional fluctuations in a $3D$ metal, as in Ref.\cite{indranil}, a full portion of the Fermi surface is hot, even at zero temperature. In that case $V_{h} $ and $V_{c}$ can be taken as  constants of the temperature. The formulae (\ref{eqn8}) is typical of electrical transport around a SDW QCP. It can be understood in the following way.  At zero temperature, the resistivity saturates to the value $ L_{11} = \pi v_F^2 \rho_0^* V_F \tau_0 /2 $. At very low temperature  for which $ \tau^{-1}_{hot}  \ll \tau_0^{-1}$, the correction to the residual resistivity acquires an anomalous exponent  $ L_{11}= \pi v_F^2 \rho_0^* V_h  \tau_h^{-1} \tau_0^2 /2$. Note that although this exponent is universal, its regime of validity can be quite small, since it requires that  $ T \leq \left ( \tau_0^{-1} \right )^{2/(d-2)}$. A good order of magnitude for the validity of this regime is that the variation of the resistivity $ \rho - \rho_0$ (or of the conductivity) over which this regime is observed must be of the same order of magnitude as $\rho_0$ itself. At even higher temperature, the resistivity is short circuited by the conduction electrons, leading to a typical form $L_{11} = \pi v_F^2 \rho_0^* V_F \tau_c /2$. These results are described in details in Ref. \cite{achim}.

 Let's now treat the  off-diagonal correlation function between the heat current and the electric current. Here too, we have two contributions, one from the hot part of the Fermi surface and one from the cold part.  From Eqn. \ref{eqn2} we see that $L_{12}$ is odd in frequency.  For the contribution not to vanish, some asymmetry has to be introduced either in the summation over ${\bf p}$  via an asymmetry in the density of states or in the summation over $\omega$ via an asymmetry in the scattering times.  For this purpose we make the phenomenological assumptions for both the impurity scattering time and the scattering time of the electron over the quantum critical  modes.
 $ \tau^{-1}_{imp } = \tau_0^{-1} + \omega \rho^\prime_0 A_{imp} $ and $\tau^{-1}_{dyn} = \tau^{-1}_{h/c} (1+  \tau_A  \  \omega ) $ where $ \tau_A$ represents  the asymmetric part of the scattering rate; it has the dimension of a lifetime. As for $L_{11}$ we find two contributions to $L_{12}$ coming from the hot and the cold regions of the Fermi surface.   \beq \label{eqn9} \begin{array}{c}
 \ds { L_{12}= L_{12}^{h} + L _{12}^{c}}  \ ; \\
 \\
  \ds {  L_{12}^{h}   =    \frac { \pi v_F^2}{2}\  T^2  \rho^{* \prime}_0  \frac {V_{h} }{  \tau_0^{-1}  + \tau^{-1}_{h} } \times} \\
     \ds {  \left[ \frac {1}{Z^{h}_\omega (T) } - \frac{\rho_0 A_{imp}}{ \tau_0^{-1}  + \tau^{-1}_{h} } + \frac{\rho_0}{\rho^\prime_0} \frac{ \tau_A \tau^{-1}_{h}  } { \tau_0^{-1}  + \tau^{-1}_{h}  }  \right ]   \  , }  \\
      \\
      \ds{  L_{12}^{c}   =    \frac { \pi v_F^2}{2}\  T^2  \rho^{* \prime}_0  \frac {V_{c} }{  \tau_0^{-1}  + \tau^{-1}_{c} }  \times } \\
    \ds { \left[ \frac {1}{Z^{c}_\omega (T) } - \frac{\rho_0 A_{imp}}{ \tau_0^{-1}  + \tau^{-1}_{c} } + \frac{\rho_0}{\rho^\prime_0} \frac{ \tau_A \tau^{-1}_{c}  } { \tau_0^{-1}  + \tau^{-1}_{c}  }  \right ] \ , }\\
    \\
      \end{array} \label{eqn3}   \eeq  with $ \tau^{-1}_{h}= Im \Sigma_c^{h} $ is the inverse scattering time in the hot regions of the Fermi surface while $ \tau^{-1}_{c}= Im \Sigma_c^{c} $ is the inverse scattering time in the cold regions. From (\ref{eqn3}) we can see  that  in the QC regime, the  contribution from the hot lines  is dominated by $1/ Z_\omega $   since   this quantity diverges as $T^{(d-z)/z}$.
    Considering that  in the quantum critical regime $L_{11}$ saturates in Eqn. (\ref{eqn8})  we get the following asymptotic form in the QC regime
    \beq \label{eqn10}
    \left | \frac{S -S_0}{T} \right | \simeq \frac{\rho^{* \prime} \ V_{h} }{\rho^*  \ V_F }  \ \frac{1}{Z_\omega^h}  \ , \eeq where $S_0$ is the saturation value of the Seebeck coefficient at zero temperature.  
      This result is  quite remarkable since it shows that for all configurations of the hot lines, the  correction to the thermopower divided by the  temperature  tracks the  variation of the Sommerfeld coefficient.  Indeed, when  the hot region has finite width, $S_0 \rightarrow \infty$ and  $ S(T) \sim 1/ Z_\omega $. When the hot regions  have the shape of a line or a point $ V_h \sim \sqrt{T}$  and $1/Z_\omega^h \sim T^{(d-3)/2}$ so that the product tracks the Sommerfeld coefficient. This result was  obtained in \cite{indranil} in the case of two dimensional fluctuations in a $3 D$ metal [however in this case the thermopower diverges at the QCP since the hot region has finite width, which is not the case when the hot region has the shape of a line or a point]. It is quite remarkable that  it generalizes to  all cases.

      Let's examine in more details the zero temperature regime around the QCP.  In the saturation regime, $ L_{12}$ is dominated by the two first terms in the brackets, both in the hot and the cold regions.  The form of $A_{imp}$ is taken from \cite{miyake1} $ A_{imp} = \tau_0^{-1} /\rho_0 \ \left ( 1 - ( \pi \rho_0 Z_0 U)^2 \right ) / \left ( 1 + ( \pi \rho_0 Z_0 U)^2 \right ) $ with  U the scattering potential of the impurities.  Within these notations we find that in the very low temperature regime
      \beq  \frac{S_0}{T} = \frac{\rho^\prime_0}{ e \rho_0} \ \left [ \frac{1}{Z_0} + \frac{1 - ( \pi \rho_0 Z_0 U)^2}{1 + ( \pi \rho_0 Z_0 U)^2} \right ] \label{eqn4} \eeq We see that typically, the sign of $S/T$  at low temperatures is determined by the sign of the derivative of the quasiparticle density of states at the Fermi level. For a typical heavy fermion compound, the hybridization between the f and c electrons lead to the following density of states $ \rho^*( \eps) \sim 2 \rho_0 D/| {\tilde \eps}_f - \eps | $ where $\rho_0 $ is the conduction electron density of states, $D$ is their bandwidth and ${\tilde \eps_f} = \eps_f - \Sigma_f ( 0) $ is the potential of the re-normalized f-levels \cite{miyake1}. We understand as well from Eqn.(\ref{eqn4})  that  for the Yb-based compounds the $S/T$ is negative since the Kondo resonance, described here by the f-level, lies below the Fermi energy (the Yb atom having 13 f-electrons, the shell is almost full) whereas the Ce-based compounds have a positive $S/T$ since the f-level lies above the Fermi energy (the Ce atom has 1 f-electron, so that the shell is almost empty).
      In the vicinity of the QCP,  the residual $ | S / T |  $  is dominated by $ 1 / Z_0 $ and we expect  it to  have a  maximum at the QCP.

\begin{figure}[htbp]
\begin{center}
\includegraphics[width=9cm]{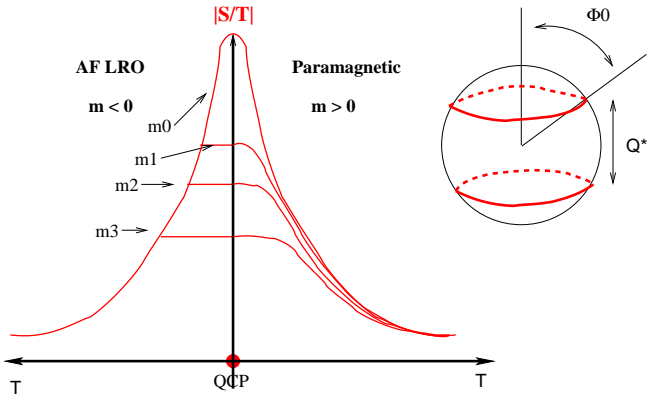}
\caption{Generic form of the Seebeck coefficient for the SDW theory in the case where the hot region has the shape of a line or a point. A saturation is present at the QCP, which is not the case for 2 D modes in a 3 D metal,  as described in Ref.\cite{indranil}.
The left side of the phase diagram corresponds to the AF phase  while the right side corresponds to the paramagnetic phase. The various curves  correspond to different value of the bare mass, which describes the proximity to the QCP.  At the QCP $m=0$, in the AF phase $m<0$ while in the paramagnetic one $m>0$. We observe a divergence of  $ | S / T |$  at the QCP, which is a generic feature independent on the  presence of the hot lines. Note that in the AF phase, the saturation is a little bit more abrupt than in the  paramagnetic phase. Apart from this  small  asymmetry feature, the phase diagram has a rough symmetric  character, typical of the SDW phase transition. }
\label{fig4}
\end{center}
\end{figure}

      The topology of the Seebeck coefficient  in the vicinity of a SDW QCP is summarized in Fig. \ref{fig4}.   The most noticeable fact about it is that $| S / T |$ is symmetric around the QCP. It  diverges at the QCP  in $ d=2$ and  increases  like  $ T^{3/2} $ and then saturates  in $D=3$ (see Table \ref{table1}). As a universal feature,  the variation of  $ |S/T|$ with temperature follows the one of the Sommerfeld coefficient $ \gamma$.

\subsection{In the vicinity of the Kondo Breakdown}

\begin{figure}[htbp]
\begin{center}
\includegraphics[width=8cm]{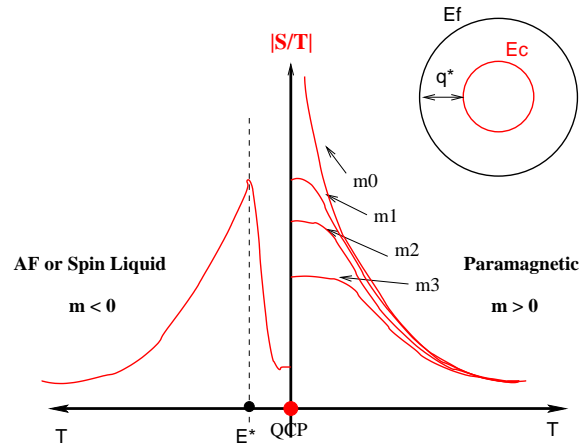}
\caption{ The form of the Seebeck coefficient divided by the temperature around the Kondo Breakdown QCP. The diagram shows a pronounced asymmetry between the heavy Fermi liquid phase and the spin liquid phase \cite{note8}. $| S / T |$ shows a brutal drop at $T=0$ when going form the Fermi liquid phase to the localized phase, which reveals the re-configuration of the Fermi surface at the QCP.  The scale $E^*$ signals a brutal change from the QC to the spin liquid regime. $S/T$ seems to be  a very good experimental probe to detect the multi-scale character of the Kondo Breakdown QCP. }
\label{fig5}
\end{center}
\end{figure}

The main difference between the SDW scenario and the Kondo Breakdown resides in the localization of the f-electrons, in the spin liquid side of the transition.  In the Kondo Breakdown scenario, the quenching of the entropy is  done through the formation of the spin liquid. We can say that the spins become entangled with one another, due to the presence of either the geometric frustration, or the frustration generated by the RKKY interactions. In the spin liquid phase the f-electrons are not entropic carriers anymore, and their contribution to the thermopower is negligible around the QCP.  One can see in figure \ref{fig5} that around the Kondo  Breakdown QCP, $|S/T|$ shows a pronounced asymmetry. On the right-hand side of the phase diagram, which corresponds to the formation of the heavy Fermi liquid,   $|S/T|$  shows the same generic structure as in the SDW case. The main response is carried by the conduction electrons,  and the scattering through the QC modes is dominant.  From Eqn.(\ref{eqn3}),  $1/Z_\omega$ is given by the scattering through the critical bosonic modes (here corresponding  to the condensation of the holons which form the heavy  quasi-particle). Away for the QCP, we observe an increase of $|S/T|$, followed by a saturation at lower temperatures. As we come closer to the QCP, the value of the saturation increases, until it diverges at the QCP.

A crucial difference with the SDW  scenario is that the critical modes have a dynamical exponent $z=3$ and not $z=2$.  As a consequence  $ |S/T |$  diverges now in dimension $d \leq 3$, with respectively a sub-logarithmic exponent  $T^{-1/3}$ in $d=2$ and a logarithmic variation $Log (T/T_0)$ in $d=3$.  We believe it is possible to detect experimentally  the difference between   the $ z=3 $ regime of the Kondo Breakdown and the $z=2$ regime of the SDW scenario, as will be developed in the next sections.
 Hence   in the QC regime, the variation of $|S/T|$ with the temperature tracks the one of the Sommerfeld coefficient [note that the whole Fermi surface is hot in this case], with
 \bea \label{eqn11}
    \left | \frac{S}{T} \right | &  \simeq &  \frac{\rho^{\prime}  }{\rho  } \ \frac{1}{Z_\omega}   \nn  \ , \\
     & \simeq  &  T^{(d-3)/3 }  \  .
     \eea

The most interesting observation concerns the left hand side of
the phase diagram, where the f-electrons have localized. As we
said, they cannot participate anymore in carrying the entropy,
which leads to a dramatic discrepancy from the SDW phase diagram.
In this part of the phase diagram, if the AF order is present,  a 
signature milder than for the SDW should be observed in $| S / T |$. The change of $| S
/ T |$ as  we pass through the AF transition is typical of  the proportion of itinerant versus localized
character of the magnetic order. If the magnetic order comes mainly from localized electrons, the jump in the thermopower
coefficient at the transition should be mild, due only to the indirect 
opening of a gap in the critical modes, as a consequence of the formation of the order.
On the other hand, it the magnetism is due to the formation of wave from itinerant electrons, we can expect that the
response in the thermopower will be significant.

The thermopower is very sensitive to the scale $E^*$ of the Kondo Breakdown. The scale $E^*$ is the energy below which the mismatch between the conduction electrons and the spinons Fermi surface becomes noticeable.  Above this energy scale the  Seebeck coefficient follows the QC regime; it is dominated by the scattering of the conduction electrons through the QC modes (here the holons). Below $E^*$,  the propagator of the QC modes is gapped, it means that the effective hybridization between the conduction electrons and the spinons  vanishes, and the Seebeck coefficient is dominated by the residual Fermi  liquid contribution. This results in a dramatic drop of $| S / T |$ below $E^*$, since  when the temperature is decreased, the conduction electron's scattering changes brutally from the  QC to the Fermi liquid regime.The two regimes  are physically  disconnected from each other.  This brutal drop at $E^*$ is similar to the  brutal decrease expected in the $T=0$ limit when
we cross the QCP from the heavy Fermi liquid side towards the
spin liquid side.  As seen in Figure \ref{fig4} the reconfiguration of the Fermi surface  is revealed by a brutal drop of $|S/T|$ when going from the Fermi liquid to the spin liquid phase. Whether the Seebeck coefficient changes sign or not will depend on the details of the  conduction scattering  in the spin liquid phase and will vary form compound to compound. A strong signature of the Fermi surface reconfiguration should be observed in the localized phase.

 We keep in mind here that  a weakness of  the theoretical treatment of the Kondo breakdown model reside in the fact that the localized degrees of freedom are badly taken into account at the present stage of the theory.   However the Seebeck  coefficient is precisely less sensitive to those degrees of freedom, since they don't carry entropy. That's why it is maybe a decisive test for revealing the scale $E^*$ and thus differentiate between the SDW and  the Kondo Breakdown QCPs.

 There is a potential issue of  whether the scale $E^*$ can be masked by the occurence of AF order.
 To be  more precise, insensitivity of the Seebeck coefficient on
the AF order is observed near the QCP, if the energy scale
$E^{*}$  is larger than the
N\'eel temperature $T_{N}$. Away from the QCP in the AF side, in the case where
$T_{N} > E^{*}$ , the situation is somewhat
complicated. The whole "Fermi surface" of spinons can be gapped below
the N\'eel temperature, and holon excitations will become already
suppressed even above $E^{*}$ . In that case, the abrupt drop in the Seebeck coefficient might
occur from $T_{N}$ instead of $E^{*}$. This feature is also
completely different from the SDW scenario. If cold regions are still present at the QCP, there will
the measurable signature in the Seebeck coefficient at the N\'eel temperature should be much milder than the abrupt dropping at
$E^{*}$ which occurs in the Kondo breakdown model. 

\section{A small survey of four compounds}

\subsection{ YbRh$_2$Si$_2$ }

 The  most recent results of the thermopower in YbRh$_2$Si$_2$, driven to the QCP via the application of a small magnetic field,  comes form Ref.\cite{hartmann}. For an applied magnetic field $ B \leq 65 m T$ ,  a negative Seebeck coefficient $S<0$   is found, in good agreement with other measurements for Yb compounds \cite{Yb-heavyfermions1,Yb-heavyfermions2,Yb-heavyfermions3,Yb-heavyfermions4}. A logarithmic increase  $ - S/T \simeq - \log ( T/{\tilde T}) $ with ${\tilde T}=  3K $ is observed in the QC regime, which is defined for this compound as the regime for which $ B= 65 mT$ and $T \leq 25 K $.  This logarithmic law is observed above  a temperature $T_{max} = 0.1 K $.    Below $T_{max}$ , $ - S / T $ drops abruptly to reach a very low value.  In Ref.\cite{hartmann} a change of sign is associated with this abrupt drop of the Seebeck coefficient, and it is argued that at $B=0$, in the left hand side of the phase diagram, the  Seebeck acquired a positive value.  It is to be noticed that at the QCP, above $T_{max}$ the variation of $-S/T$ follows the variation of the Sommerfeld coefficient. This behavior changes below $T_{max}$  with the sudden decrease of $-S/T$ whereas the Sommerfeld coefficient shows an upturn as the temperature is lowered.
 Another very anomalous property is that no sign of the magnetic phase transition is observed in  $- S/T$; the only temperature scale observed in this part of the phase diagram being $T_{max}$, the scale of the abrupt decrease.

 This body of results can be simply interpreted with the Kondo Breakdown theory. The abrupt change in $-S/T$ is attributed to the reconstruction  of the Fermi surface around the Kondo breakdown QCP.  The logarithmic increase is naturally interpreted with a $z=3$ QC regime, in $d=3$, which is precisely the prediction of the Kondo breakdown theory.  The evidence for a $z=3$ QC regime in this compound, is supported as well from the  logarithmic variation of the Sommerfeld coefficient with temperature and the variation of the Gr\"uneysen ratio like $T^{-2/3}$\cite{KS-gruneysen}.
  The phase for $B<0$ can be interpreted as the phase where the f-electrons localize,  and  $T_{max}$ is within the KB theory the  scale $E^*$  below which the conduction electrons become insensitive to the scattering through the QC modes.   The discrepancy below  $E^*$ between the variation of  $- S/T$ and the  Sommerfeld coefficient $\gamma$ seems to indicate that the upturn in $\gamma $ is due to the presence of localized moments, which don't participate in the transport of entropy.

  In Figure \ref{fig2} we have put YbRh$_2$Si$_2$ exactly at the crossing point between  the AF line and the Kondo Breakdown line of QCPs.  Within the interpretation of our theory, it is just an accidental fact, but this interpretation is under debate within the community. It would be very interesting to get an experimental insight on what happens when one goes a bit away from this  intersection.  Recently, YbRh$_2$ Si$_2$ has been doped with a few $\%$ of  Ir and Co.  The Ir doping pushes the compound outside the AF phase while the Co doping pushes it inside the AF phase.  It would be of the greatest interest to measure the thermopower  in the case of Ir and Co  doping.  Within the Kondo breakdown theory we expect that the features of the  pure compound will be reproduces with no major changes as soon as $E^* \geq T_N$.
  The presence or absence of magnetic order should not affect   in a major way the location of the scale $E^*$ ( called $T_{max}$ in Ref\cite{hartmann}).
For doping with Co, we  will probably be in the situation where $E^{*} > T_{N}$,   in which case we can expect some changes to start at $T_N$ rather than $E^*$ but those will be of small magnitude compared to the drop at  $E^*$.
  The abrupt drop of the thermopower at the QCP can indicate in a precise manner the location of the  Kondo breakdown QCP, and as such the measurement of  thermoelectric effects is decisive in corroborating or invalidating the results of Ref. \cite{YRS-ir-co-doped}.  Of particular interest is the confirmation of a QCP under the AF dome in the Co-doped YbRh$_2$Si$_2$. Lastly, some measurements under hydrostatic pressure are necessary to validate the whole picture\cite{knoebel}.

   In this compound   the  application of pressure or doping has important consequences on the structure of the magnetism. It mainly affects the amount of frustration and the dimensionality of the magnetic order. The thermopower however, occurs to be mostly insensitive to  the details of the magnetic order, as soon as  it is of localized character.  That's what makes it such an attractive experimental probe to test the Kondo breakdown scenario.

 \subsection{Ce(Mn)In$_5$}

  In this series of compounds, the only study of the thermopower close to a QCP concerns CeCoIn$_5$ \cite{kamran2}. The compound is superconducting at $T=0$, and driven to  a QCP around $H_{c2} \sim 5.5 T $.   Around this field driven QCP the thermoelectric properties have been thoroughly investigated in Ref.\cite{kamran2}. The results certainly don't show any strong re-structuration of the Fermi surface at the QCP.  When the applied field crosses  the QCP,  a small increase of $ S/T$ is observed. The Seebeck coefficient is positive in the whole phase diagram. These results suggest that the QCP is not a Kondo Breakdown, but maybe in the universality class of an AF SDW, or in   a third universality class, that could appear in the presence of a strong magnetic field. In particular, the fact that the ratio $ q = ( S N_A e)/( \gamma T ) $ departs from unity at the QCP, which is well reproduced by the SDW scenario \cite{miyake1}.

  In Figure \ref{fig2} we have put CeCoIn$_5$ on the right side of the Kondo Breakdown QCP (the SC phase has not been represented here). It is possible that under a magnetic field, CeCoIn$_5$ is driven towards a QCP associated with short range AFM.

  For this series of compound, the best change to find the Kondo breakdown QCP is  around CeRhIn$_5$.  This compound is an AFM at low temperatures. With the application of pressure, it is driven towards a phase transition around $1.75 G Pa$ where the Fermi surface re-configures \cite{tuson-nature}.
 It our belief that this QCP is associated with the Kondo Breakdown, with a quasi-two dimensional nature of the QC fluctuations\cite{ronning-transport}. If it is the right hypothesis, the measurement of the thermopower  under pressure, around the point where the Fermi surface re-configures should show  a dramatic change, with  $S/T$ dropping off from the heavy Fermi liquid at high pressure to the local f-electron phase at low pressure.  An interesting point  to investigate here is whether the magnetic order at zero pressure is of itinerant or localized character or maybe both at the same time\cite{magnetism-115}. If the magnetic moments are  fully localized character, no strong signature of $T_N$ shall be observed in $S/T$, whereas, the scale $E^*$ precursor of the reconfiguration of the Fermi surface, shall be observed instead. On the other hand, if the magnetic order is of itinerant character, a signature of $T_N$ comparable to the one observed in the specific heat measurement shall be observed.
 It is possible to apply magnetic field as well, leading to a line of QCP with the magnetic field. If it is possible, it would be very interesting to measure the thermopower  close to this cline of QCP.  When following the line of QCPs, a cross over towards SDW type scenario might be observed, similar to the one found in CeCoIn$_5$.

 In this compound as well, the study of  thermoelectric  properties would be a decisive experiment in order to elucidate the nature of the QCP under pressure.

\subsection{CeCu$_{6-x}$Au$_{x}$}

This compound is one of the first where  the presence of a QCP was detected\cite{MatthiasHilbert}. When this compound is doped with $  0.1 \% $ of Au one reaches  an AF QCP.  In the QC regime the specific heat was shown to vary logarithmically with temperature, while the resistivity is linear in T.  Neutron scattering experiments \cite{schroeder} have revealed that the dynamic spin susceptibility has a pronounced two dimensional character, and shows anomalous exponents for a wide range of ${\bf q}$-vectors in the Brillouin zone.

 Two theories are in competition for this compound.  In Ref.\cite{rosch} it was argued that the QCP is very anisotropic, and its nature  is one of a  two dimensional SDW in a three dimensional metal. This theory reproduces the linear corrections to the resistivity, as shown in Table \ref{table1}.  Since  the  chemical doping with Au introduces a substantial amount of  disorder, it  is conceivable that the linear resistivity observed in this compound is  due to a wide tail correction to the residual resistivity, in the  $d=2$ SDW scenario.   In $d=2$ the  SDW scenario is believed not to produce any anomalous exponent when the  electrons have been integrated out of the partition function\cite{florens}. It has been argued, however  that it is no so when  the electrons are treated self- consistently with  the quantum critical  modes. In that case, anomalous exponents have been predicted for the staggered  dynamical spin susceptibility, in the QC regime \cite{chub3}. The observed  linear variation of the transition temperature $T_N$ with the doping x  corresponds as well to the two dimensional character of the  QC  fluctuations  within the SDW theory.

  Another theory has been proposed to explain the anomalous properties, called the locally quantum critical scenario\cite{qimiao}. This theory assumes that the bosonic modes have two dimensional character, and then using extended  Dynamical Mean Field Theory (e-DMFT) it is argued in this work that a local mode emerges at the QCP, leading to a reconfiguration of the Fermi surface.

  As a result the right theory for this compound is still very mysterious; is it a SDW with two dimensional character, or a more unconventional locally quantum critical scenario ?

  The only study of  thermoelectric effects is a short note where it is shown that the  thermopower diverges at the QCP \cite{lohneysen}.  It  is not clear what is the exponent of this divergence.  It would be very interesting to measure the thermopower in the AF magnetic phase of this compound. If the Fermi surface is re-configured around the QCP, an abrupt change of  $S/T$ is predicted. On the other hand, if the  two dimensional SDW scenario is the right answer, one should observed  in $S/T$ a  consequent signature of  the Néel temperature
  or the order of the one observed in the Sommerfeld coefficient $\gamma$. Moreover, the phase diagram in that case will be show broadly symmetric features between the ordered phase and the paramagnetic phase, with no abrupt change of $S/T$ at the QCP.
  In Figure \ref{fig2} we have placed CeCu$_{6-x}$Au$_x$ at the proximity of the AF order, indicating that in our opinion, the most likely theory   to apply here is the one of Refs\cite{rosch,indranil}. However, itis only an opinion, and we don't have enough substantial scientific arguments to corroborate it at the moment.
  Here again, the measurement of the thermoelectric effects appears to be  a decisive experiment.

\subsection{URu$_2$Si$_2$}

This compound  constitutes one of the most enduring mysteries in the field of strongly correlated electrons.   A very well defined phase transition carrying more than $ 40 \%$ of the free electron's entropy  occurs below $T_0= 17 K$. Despite  almost twenty five years of experimental investigations, the mystery concerning the nature of this ``hidden order'' remains unsolved \cite{urusi}.
Several theoretical  proposals have been made, including some exotic  short range anti-ferromagnetism \cite{premi}, a Lifshitz transition \cite{varma},  a charge density wave scenario \cite{balatsky} and a scenario where the hidden order has a localized character \cite{kiss,mineev}. Recent experiments \cite{andres}  have revived the bad structure studies. A debate exists on the nature of the two f-electrons in the U-f$^2$ atom, whether the localized picture is correct \cite{haule} or whether the itinerant one\cite{openeer} is the correct picture.
Very interesting experiments under pressure show  a long range AF (LRAF) order occurs at  the pressure of $0.5 GPa$, and the transition form the hidden order to the LRAF order is of the first order\cite{hassinger}.  Neutron scattering experiments show two types of excitations, one at ${\bf Q}_{AF} = (0,0,1)$, which becomes static and long range under pressure in the AF phase\cite{villaume}, and another excitation at ${\bf Q}_0 = ( 1,0,0) $, which  is present only in the hidden order phase\cite{wiebe}.
The superconducting phase is as well of very unconventional nature \cite{machida}
The study of thermoelectric effects in URu$_2$Si$_2$ is complicated by the fact that it is a compensated metal.  Below $T_0$ it  has  been established that the number of carriers drops considerably, leading to  the physics of very low density of electrons \cite{kamran4.1,kamran4.2,kamran4.3}.

With all these observations in mind, it might look surprising to test URu$_2$Si$_2$ as a potential candidate  for the proximity to a QCP. A recent study of thermoelectric effects on this compounds might change this perspective. In the paper Ref\cite{kamran3}, a thorough study of both resistivity and thermopower has been conducted under magnetic field. This study first confirmed the strong anisotropy of this compound.
The anisotropic nature of this compound was known for a long time,  with the observation of anisotropy factor of 3 to 5 in the resistivity\cite{anis-res}, the magnetic susceptibility\cite{anis-chi} and the critical field\cite{anis-crit}. The de Haas van Alphen study for this compound captures  only a mild anisotropy in the three Fermi surfaces observed. Very interestingly  Ref\cite{kamran3} reveals  a significant anisotropy in the inelastic scattering of the normal phase.  When a magnetic field of $12  T$ is applied, anomalous scattering is observed in the electrical resistivity of the basal plane, which is then linear in temperature, while the c-axis resistivity remains Fermi liquid like down to very low temperatures.  The Seebeck coefficient divided by temperature shows as well a very anomalous behavior. In the basal plane it departs from the constant value predicted by the Fermi liquid theory to finally change sign  for fields larger than $12 T $, at temperatures lower than $T_{change}= 0.8 K$. In the c-axis, the signal remains Fermi liquid like, with a well defined saturation for all fields considered. This anisotropic situation is very reminiscent of the case of CeCoIn$_5$,  for which as well an inelastic transport time has been revealed\cite{paglione}.

This body of observations motivates us to suggest that
URu$_2$Si$_2$ might be in the proximity of an anisotropic Kondo
Breakdown QCP. The  anisotropic scattering is cut-off below the
scale $E^*$   characterizing the Kondo Breakdown theory. In the
present case $E^*$ would be anisotropic, with a small value of the
order of $E^*_{ab} \sim 0.8 K $  in the basal plane and with a
much bigger value of $E^*_c$ in the c-axis direction.  A detailed
exposition of this proposal will be published
elsewhere\cite{note3}. At the moment we would like to suggest that
it would be extremely interesting to test this idea by exploring
the thermopower on the whole pressure phase diagram. If the AF
order is of localized nature, no significant  entropy transport
should be associated with the occurrence of the AF phase, and  the
signature in the Seebeck coefficient should be minor
near the QCP.
On the other hand, if the AF order is of itinerant character, a
strong signature in $S/T$ is to be observed. Moreover, if the
compound is sitting at the proximity of the Kondo Breakdown QCP,
one expects to see some evidence of the scale $E^*$ at other
pressures in the phase diagram, and especially in the LRAF phase.
In the Figure \ref{fig2} we have placed  URu$_2$Si$_2$ in the
proximity, but still a little bit away from the Kondo Breakdown
QCP. For this compound again, thermoelectric studies under
pressure could be decisive  to unveil the mysterious nature of the
hidden order.

\section{Conclusion}

 The aim of this discussion paper is principally to encourage new experiments using the thermopower as a testing probe for
 discriminating the nature of QCPs in heavy fermions.
 It turns out that the two main classes  of QCPs in heavy fermions have very different signatures in terms of the Seebeck coefficient.
 The SDW scenario   has  Seebeck coefficient  with a good degree of  symmetry  around the QCP,
  between the ordered phase and the paramagnetic phase.
 On the other hand,  for the Kondo Breakdown QCP, the Seebeck coefficient  shows a pronounced
  asymmetry around the phase diagram,  dropping out in the  Kondo broken phase,
  since the f-electrons are not available anymore to carry the entropy and the quantum critical scattering
  of the conduction electrons is gapped below   an energy scale $E^*$.

  The Seebeck coefficient can be used as a very sensitive probe to detect whether the  magnetism is of localized or itinerant character. In the case of itinerant magnetism,
  $S/T$  is qualitatively tracking the variation of heat at the magnetic transition.  For magnetism emerging from localized moments, the  specific heat is expected to
   be one order of magnitude more sensitive to the phase transition than  $S/T$,  since in that case the  localized f-electrons don't participate to the heat transport
   whereas their entropy is locally quenched by the apparition of the order.

   Lastly, the temperature dependence of $S/T$ in the QC regime is tracking the variation in temperature of the Sommerfeld coefficient, which enables to make the distinction between
   different classes of  QCP, with dynamical exponent $z=2$ or $z=3$.

    It is our belief that new experiments within this technique, especially under pressure, can shed light on the nature of the various QCPs of heavy fermion compounds.
    
    We thank I.Paul for interesting discussions and suggestions and Q. Si, M. Vojta  and S. Hartmann for comments and suggestions on the manuscript.
     C.P. thanks the Aspen Center for Physics where the idea of writing this paper emerged as well as ICAM, for Senior Travel Fellowship.

\appendix*

\section{Appendix: derivation of eqn.(\ref{eqn10})}
In this Appendix we derive the equations leading to (\ref{eqn8}), (\ref{eqn9}) and (\ref{eqn10}).
\subsection{The current-current correlation function}

We start with
\[ \begin{array}{cl} &  L_{11} = \sum_{\bf p}  v_{\bf p}^2  \int_{- \infty}^{\infty} d \omega \left(- \frac{ \partial f}{\partial \omega} \right )  A^2({\bf p}, \omega)   \\
\\
  \mbox{with} & \ds{  A( {\bf p}, \omega) = \frac{\tau^{-1}(\omega) } { ( \omega/Z_\omega -\eps_{\bf p}  )^2 +  \tau^{-2}(\omega) } }   \\
  \\
  \mbox{and} & \tau^{-1} (\omega) : = \tau^{-1}_0  + \tau^{-1}_{dyn}( \omega)  \ . \end{array}  \]
  The value of $\tau^{-1}_0  $ and $ \tau^{-1}_{dyn}$ are given in the text in Eqns.(\ref{tau}) and (\ref{taubis}). The  scattering time $\tau$ is considered here as valid respectively in the ``hot'' and ``cold'' region, and the subscript has been omitted.
   Using $ \sum_{\bf p }  =  \int_{-D}^D \rho ( \eps) d \eps $ and noticing that the wave vector $v_{\bf p } $ is pinned at the Fermi surface, we get

\beq \begin{array}{l}
 L_{11} = v_F^2  \rho_0 \int_{-\infty}^\infty d \omega \left(- \frac{ \partial f}{\partial \omega} \right ) \int_{-D}^D  d \eps \ \left ( \frac{ \tau^{-1} (\omega )}{\eps^2 + \tau^{-2}(\omega )  }  \right )^2 \ , \\
 \\
 \mbox{changing variables for  } \eps \rightarrow y \mbox{  we get}\\
 \\
 L_{11} =   v_F^2  \rho_0 \int_{-\infty}^\infty d \omega \left(- \frac{ \partial f}{\partial \omega} \right ) \tau(\omega)  \int_{-\infty}^\infty d y  \left (\frac{1}{ y^2+1} \right )^2 \ , \\
 \\
 \mbox{ and remembering that depending on the region}\\
 \mbox{ in the Fermi surface,  } \tau_{dyn} = \tau_h \mbox{  or  } \tau_{dyn} = \tau_c \ , \mbox{  we get} \\
 \\
 L_{11} = \ds{ \pi  \frac{v_F^2}{2} \left (   \frac{\rho^*_0 V_{h}}{ \tau_0^{-1} + \tau^{-1}_{h} }   +  \frac{ \rho^*_0 V_{c} }{ \tau_0^{-1} + \tau^{-1}_{c} }  \right ) \ ,} \end{array} \eeq with
 in spherical coordinates $ \rho^*(\eps ) = \frac{ p^2 d p }{ ( 2 \pi)^2  d \eps }  $, $ V_{h} =  \sin \phi_0 \Delta \phi $, $V_{c} = 2 -   \sin \phi_0 \Delta \phi$ and $ \Delta \phi  \sim \sqrt{T} $ the width of the hot  regions.

 \subsection{The heat-current correlation function}
 The heat-current correlation function is given by
 \[ \begin{array}{cl} &  L_{12} = \sum_{\bf p}  v_{\bf p}^2  \int_{- \infty}^{\infty} d \omega \left(- \frac{ \partial f}{\partial \omega} \right )  \omega A^2({\bf p}, \omega)   \\
\\
  \mbox{with} & \ds{  A( {\bf p}, \omega) = \frac{\tau^{-1}(\omega) } { ( \omega/Z_\omega -\eps_{\bf p}  )^2 +  \tau^{-2}(\omega) } }   \\
  \\
   & \tau^{-1} (\omega) = \tau^{-1}_{imp}  + \tau^{-1}_{dyn}  \ ,\\
   \\
     & \tau^{-1}_{imp}(\omega ) = \tau^{-1}_0 +  \omega \  \rho^\prime_0 \ A_{imp}( 0)  \ , \\
     \\
  \mbox{and} & \tau^{-1}_{dyn} (\omega) = \tau^{-1}_{h/c} \ ( 1 + \tau_A \ \omega )  \ .  \end{array}  \]
The definition of  $\tau^{-1}_{h/c}$ is given in the text in Eqn. (\ref{taubis}).
  The heat-current correlation function is evaluated with the following steps. Transforming $ \sum_{bf p} $ into an integral over $ \eps$ we get
  \beq \begin{array}{l}
 \ds {L_{12} = v_F^2   V_{h/c}  \int_{-D}^D d \eps \left ( \rho^*_0 + \eps \rho^{* \prime}_0  \right ) } \\
 \ds{ \times  \int_{-\infty}^\infty d \omega \left(- \frac{ \partial f}{\partial \omega} \right ) \omega \left ( \frac{  \tau^{-1} (\omega )}{( \omega/Z_\omega -\eps  )^2 + \tau^{-2}(\omega )  } \right )^2} \ , \\
 \\
 \mbox{ permuting the integrations and  changing variables  } \\
 \eps \rightarrow  \eps + \omega/Z_\omega \ , \mbox{  we get}\\
 \\
 \ds{ L_{12} = v_F^2   V_{h/c}  \int_{-\infty}^\infty d \omega \left(- \frac{ \partial f}{\partial \omega} \right ) \omega   } \\
\ds{  \times  \int_{-D-\omega/Z_\omega}^{D+\omega/Z_\omega} d \eps \left (\rho^*_0 + (\eps +\omega/Z_\omega )   \rho^{* \prime}_0\right ) \left (\frac{\tau^{-1}(\omega) }{ \eps^2 + \tau^{-2} ( \omega )  } \right )^2}  \ , \\
\\
\mbox{then changing variables  } y := \eps \tau(\omega) \\
\\
\ds{ L_{12} = v_F^2   V_{h/c}  \int_{-\infty}^\infty d \omega \left(- \frac{ \partial f}{\partial \omega} \right ) \omega \tau(\omega)  } \\
\ds{  \times  \int_{- \infty}^{\infty} d y \left (\rho^*_0 + (y \  \tau^{-1} (\omega ) +\omega/Z_\omega )  \  \rho^{* \prime}_0\right ) \left (\frac{1 }{ y ^2 + 1   } \right )^2}  \ , \\
\\
\mbox{the term linear in y vanishes and the contribution}\\
\mbox{ in front of  }  \rho^*_0 \mbox{  reads}\\
\\
\ds{ L_{12}^a = \frac{ \pi v_F^2}{2}   V_{h/c}\ \rho^*_0  \int_{-\infty}^\infty d \omega \left(- \frac{ \partial f}{\partial \omega} \right ) \omega \tau(\omega)  } \\
\\
\mbox{using the definition of  } \tau(\omega ) \mbox{  we have}
\\
 \ds{ L_{12}^a = \frac{ \pi v_F^2 }{2}   V_{h/c} \  \rho^*_0 ( - T^2)  \frac{\rho^\prime_0 A_{imp}(0) + \tau^{-1}_{h/c} \  \tau_A }{( \tau_0^{-1} + \tau^{-1}_{h/c} )^2 }  }  \\
 \\
 \mbox{the contribution in  } L_{12} \mbox{  in front of  } \rho^{* \prime}_0 \mbox{  reads}\\
 \\
 \ds{ L_{12}^b = \frac{ \pi v_F^2}{2}   V_{h/c}\ \rho^{* \prime}_0  \int_{-\infty}^\infty d \omega \left(- \frac{ \partial f}{\partial \omega} \right ) \omega^2 \tau(\omega) Z^{-1}_\omega  }\\
 \\
 \mbox{which gives}\\
 \\
  \ds { L_{12}^b =  \frac{\pi v_F^2 }{2}   V_{h/c} \  \rho^{* \prime}_0 ( T^2) \frac{1}{\tau_0^{-1} + \tau^{-1}_{h/c}} \ \left . \frac{1}{Z_\omega} \right |_{\omega=T}  } \\
  \\
  \mbox{finally}
  \\
  \\
  L_{12}= L_{12}^a+ L_{12}^b  \ . \end{array} \eeq


\begin{thebibliography}{99}
\bibitem{note1} The dynamical exponent counts the number of equivalent dimensions in space that the quantum fluctuations occupy.
\bibitem{note5} The robustness of the $z=3$ QC regime of the Kondo Breakdown is still a matter of  discussion in the community. It will be shown in a subsequent publication that
this regime is robust towards the variation of the shape and the location of the two Fermi surfaces (the one of the $\chi$-spinons and the one of the c-electrons).
\bibitem{note2} The holon propagator is the propagator of the order parameter of the Kondo Breakdown.
\bibitem{notestar} I has to be  noticed that theoretically the  Fermi liquid is broken below  $E^*$ \cite{senthil1}. We mean here that those deviation from the Fermi liquid theory are not appreciable in transport properties. The thermodynamics, however might show some signature of the gauges modes.
\bibitem{note8} Note that the scale $E^*$ is supposed to be decreasing in the heavy Fermi liquid phase, as will be shown in a subsequent publication. 
This feature supports the asymmetry of the thermopower around the phase diagram.
\bibitem{note3} The authors of the present paper are currently working on this proposal.
\end{thebibliography}
\end{document}